\documentstyle[preprint,tighten,aps]{revtex}
\begin{document}
\draft
\hyphenation{Nijmegen}
\hyphenation{Rijken}
 
\title{Can the magnetic moment contribution explain the
       $\bbox{A_y}$ puzzle?}
 
\author{V.G.J.\ Stoks}
\address{TRIUMF, 4004 Wesbrook Mall, Vancouver, B.C., Canada V6T 2A3 \\
         and Physics Division, Argonne National Laboratory,
         Argonne, IL 60439, USA}
 
\date{}
\maketitle

\begin{abstract}
We evaluate the full one-photon-exchange Born amplitude for $N\!d$
scattering. We include the contributions due to the magnetic moment
of the proton or neutron, and the magnetic moment and quadrupole moment
of the deuteron. It is found that the inclusion of the magnetic-moment
interaction in the theoretical description of the $N\!d$ scattering
observables cannot resolve the long-standing $A_y$ puzzle.
\end{abstract}
\pacs{13.75.Cs, 11.80.Et, 21.45.+v}


With the advent of high-speed computers and improved algorithms it has
now become feasible to do realistic $N\!d$ scattering calculations.
The state-of-the-art calculations are those of the Bochum-Cracow
group~\cite{Wit87} in momentum space and of the Pisa group~\cite{Kie94}
in coordinate space. Results for eigenphase shifts and mixing parameters
from both methods are very close~\cite{Hub95}. Using recently constructed
high-quality $N\!N$ potential models~\cite{Sto94,Wir95,Mac96}, it is
found that almost all experimental data, both below and above breakup,
can be described very well~\cite{Glo96}. The only exceptions are the
low-energy $pd$ and $nd$ analyzing powers. Up to nucleon laboratory
energies of about 30 MeV the theoretical predictions using realistic
$N\!N$ potentials are systematically lower by about 30\% than the
experimental measurements. Inclusion of 3N forces has not improved the
situation~\cite{Wit94}. This discrepancy has become known in the
literature as the $A_y$ puzzle.

An attempt to resolve this puzzle has been to modify the $^3P_J$ $N\!N$
partial waves. It is found that only a slight modification is required
to be able to describe both $N\!N$ and $N\!d$ observables accurately. 
The reason is that the $N\!d$ analyzing power $A_y$ is very sensitive
to the $N\!N$ $^3P_J$ waves, whereas at the energies under consideration
there are too few $N\!N$ analyzing power data to put a real severe
restriction on the individual $^3P_J$ waves. In their first attempt to
modify the $^3P_J$ waves, Wita{\l}a and Gl\"ockle~\cite{Wit91} indeed
found a set of phase shifts which allowed for a satisfactory description
of both 2N and 3N observables simultaneously.
However, the resulting phase shifts exhibit a large breaking of charge
independence and charge symmetry; too large, in fact, to be consistent
with theoretical predictions based on meson-exchange models of the
$N\!N$ interaction.

Keeping the amount of charge-independence breaking at a more realistic
level, it was recently shown~\cite{Tor97} that it is still possible to
find a set of $^3P_J$ phase shifts that leaves the description of the 2N
data basically untouched and at the same time largely reduces the
discrepancy in the description of the $nd$ analyzing power at $E_n=3$ MeV.
Different energies and $pd$ scattering have not been analyzed yet.
In this approach each individual $^3P_J$ phase shift (from a potential
model or a partial-wave analysis) is modified with a different
multiplicative factor $\lambda_J$. It is found that $\lambda_0,\lambda_1<1$
and $\lambda_2>1$, and so the $\lambda_J$ factors cannot be associated
with some modification to one-pion exchange, but have to be due to an
increase in strength of the short-range interaction.
However, in $N\!d$ scattering one probes the $N\!N$ interaction over
an extended energy range, and so the $\lambda_J$ factors appear to be
energy dependent. It is not clear yet how these factors can be explained
within a meson-exchange model, why they only appear to be necessary in
the $^3P$ waves, and why this apparent alternative solution for the
$N\!N$ phase shifts has not been found in partial-wave analyses.
This is presently under investigation.

In this Brief Report we will investigate an alternative way to ``modify''
the $^3P_J$ waves, which does not involve any modification of the $N\!N$
nuclear interaction. The investigation is motivated by the fact that about
ten years ago we experienced a similar problem in $N\!N$ scattering.
It was found that also in that case the (at that time) current theoretical 
approach was unable to properly describe the new high-accuracy $pp$ and
$np$ analyzing power data. The reason could be traced to an incorrect
description of the spin-orbit combination of the $^3P$ waves. The proper
inclusion of the magnetic-moment interaction due to the magnetic
moments of the proton and neutron was found to completely resolve this
problem~\cite{Sto90}. Furthermore, also in this case the modification
to the $^3P$ waves is very small but crucial in describing the data
correctly. More importantly, the inclusion of the magnetic-moment
interaction in the $N\!N$ scattering mainly affects only the analyzing
power, whereas the other observables (differential cross sections,
spin correlation parameters, rotation parameters, etc.) remain basically
unaffected. We are looking for a similar solution to the $A_y$ puzzle:
the theoretical description of the $N\!d$ analyzing powers needs to be
modified, but the other observables which can already be described
very well (such as differential cross sections and tensor analyzing
powers) should remain more or less unaffected. Hence, an obvious
candidate for a solution to the $A_y$ puzzle would be to include the
effects due to the magnetic and quadrupole moments. Here we will
investigate whether this indeed provides a solution.

Since in this first investigation we only want to find out whether the
inclusion of the magnetic-moment interaction might indeed provide a
solution to the $A_y$ puzzle, at this stage we refrain from doing a
full calculation in all its complicated glory. We therefore here do
not view the deuteron as a two-body ($np$) bound state, but rather as
a single spin-1 particle with charge, magnetic moment, and quadrupole
moment. The full electromagnetic amplitude for $N\!d$ scattering is
then simply given by the Born amplitude of the one-photon exchange
between a spin-1/2 particle (the nucleon) and a spin-1 particle
(the deuteron). In case of $pd$ scattering the amplitude has to be
calculated in Coulomb distorted-wave Born approximation, which
substantially complicates the calculation. A convenient and accurate
approximation is to first calculate the plane-wave Born approximation
and then to modify the amplitude with the so-called Breit
factor~\cite{Bre55}. Again, this simplification is completely adequate
for our present purposes.

The one-photon-exchange Born amplitude for $N\!d$ scattering is given by
\begin{equation}
   M^{\gamma}_{Nd} = iej^{\mu}_{N}\,\frac{1}{8\pi\sqrt{s}t}\,
                     iej_{d\mu} = -\frac{\alpha}{2t\sqrt{s}}\,
                     j^{\mu}_{N}j_{d\mu},           \label{mndgam}
\end{equation}
where $s$ and $t$ are the standard Mandelstam variables, and $j_{N}$
and $j_{d}$ are the nucleon and deuteron currents, respectively.
The nucleon current in the various spin-1/2 projection states is
given by the $2\times2$ matrix
\begin{equation}
   \langle {\bf p}'_N,m'|j^{\mu}_{N}|{\bf p}_N,m\rangle =
    \overline{u}({\bf p}'_N,m')\left[F^N_1\gamma^{\mu}
        +\frac{iF^N_2}{2M_N}\sigma^{\mu\nu}k_{\nu}\right]
        u({\bf p}_N,m),                 \label{Jnuc}
\end{equation}
with $u({\bf p},m)$ a Dirac spinor, $k=p'_N-p_N$ the momentum transfer,
and $F^N_1$ and $F^N_2$ the nucleon charge and magnetic-moment form
factors.
The deuteron current is given by the $3\times3$ matrix
\begin{equation}
   \langle {\bf p}'_d,m'|j^{\mu}_{d}|{\bf p}_d,m\rangle =
    \varepsilon^{\ast\rho}({\bf p}'_d,m')\left\{
        (p'_d+p_d)^{\mu}\left[g_{\rho\sigma}F^d_1
        -\frac{F^d_2}{2M_d^2}k_{\rho}k_{\sigma}\right]
        +I^{\mu\nu}_{\rho\sigma}(p'_d-p_d)_{\nu}G^d_1\right\}
    \varepsilon^{\sigma}({\bf p}_d,m),         \label{Jdeut}
\end{equation}
where $I^{\mu\nu}_{\rho\sigma}$ is the infinitesimal generator of the
Lorentz transformation, and $F^d_1$, $F^d_2$, and $G^d_1$ are the deuteron
form factors. The polarization vectors $\varepsilon^{\mu}({\bf p},m)$
satisfy
\begin{eqnarray}
    && \varepsilon^{\ast}_{\mu}({\bf p},m')\varepsilon^{\mu}({\bf p},m)
    =-\delta_{m',m},   \nonumber\\
    && \sum_{m}\varepsilon^{\ast}_{\mu}({\bf p},m)\varepsilon_{\nu}({\bf p},m)
    =-g_{\mu\nu}+\frac{p_{\mu}p_{\nu}}{M^2_d},     \nonumber\\
    && p_{\mu}\varepsilon^{\mu}({\bf p},m)=0,
\end{eqnarray}
and are given, for example, in Ref.~\cite{Noz90}.
Given the above expressions, it is straightforward to evaluate the 
$6\times6$ one-photon-exchange $pd$ and $nd$ Born amplitudes.
As mentioned before, for $pd$ scattering the amplitude has to be modified
with the Breit factor:
\begin{equation}
    \frac{1}{2t} \rightarrow 
    \frac{\exp[-i\eta\ln\frac{1}{2}(1-\cos\theta)]}{2t},
\end{equation}
with $\eta$ the standard Coulomb parameter and $\theta$ the scattering
angle.

The electromagnetic amplitude is added to the nuclear $N\!d$ amplitude,
where the latter is calculated in the formalism presented by
Seyler~\cite{Sey69} using phase shifts of the Pisa group~\cite{Kie96}
at $E_{\rm lab}=1$, 2, and 3 MeV.
Of course, in a proper treatment of the electromagnetic addition the
nuclear amplitude has to be modified with square-root factors of the
electromagnetic $S$ matrix (for details in the analogous case of $N\!N$
scattering, see Ref.~\cite{Sto90}), but these corrections are rather
small and can be neglected at this stage of the calculation. In the
case of $pd$ scattering, however, the correction factors $e^{i\sigma_L}$,
with $\sigma_L$ the Coulomb phase shifts, {\it are} included.
Given the full amplitude $M^{\gamma}_{N\!d}+M^{\rm nuc}_{N\!d}$ it is
then straightforward to calculate the observables~\cite{Sey69}.

In Fig.~\ref{fig1} we show the $pd$ differential cross section, the
proton analyzing power, and the deuteron vector and tensor analyzing
powers at $E_p=3$ MeV ($E_d=6$ MeV). The experimental data are from
Shimizu {\it et al.}~\cite{Shi95}. It is seen that the inclusion of the
full electromagnetic amplitude hardly affects the description of the
observables. There is a small enhancement in the forward direction
for the proton analyzing power, but it is far too small to explain
the discrepancy with the experimental data. Furthermore, there is
hardly any effect at all on the deuteron vector analyzing power $iT_{11}$.
We believe that a more careful treatment (without the approximations
pointed out above) will not change this result.

In Fig.~\ref{fig2} we show the similar results for $nd$ scattering at
$E_n=3$ MeV ($E_d=6$ MeV), where the experimental differential cross
section data are from Schwarz {\it et al.}~\cite{Sch83} and the neutron
analyzing power data from McAninch {\it et al.}~\cite{McA94}. Again,
the effects are very small, where the effect of the inclusion of the
electromagnetic amplitude on the deuteron tensor analyzing powers
cannot be seen at all (the curves lie on top of each other). The effect
only shows up in the neutron and deuteron vector analyzing powers at
extreme forward angles. Below breakup there are no accurate data yet,
but above breakup, at $E_n=6.5$ MeV, the experimental data at forward
angles clearly exhibit the crossing to negative values~\cite{Tor91}.
This means that if one properly wants to describe the $N\!d$ data using
some theoretical model, the full electromagnetic $N\!d$ amplitude needs
to be included already at these low energies, even if its effect on the
intermediate- and large-angle data is very small. This is very similar
to the situation in $np$ scattering.

At lower energies, $E_N=1$ and 2 MeV, the results are very similar, and
so we conclude that although the inclusion of the full electromagnetic
amplitude mainly only affects the analyzing powers, the effects are
far too small to explain the $A_y$ puzzle. We do not believe that a more
careful treatment, without the approximations such as pointed out for
$pd$ scattering, or a detailed 3-body calculation (if at all feasible)
will significantly change these results.
The problem of the $A_y$ puzzle therefore still remains.

\acknowledgments
I thank prof.\ J.J.\ de Swart for encouraging me to do this
investigation, and prof.\ W.\ Tornow for providing the numerical
values of the experimental data.
This work was supported by the Natural Sciences and Engineering
Research Council of Canada and by the U.S.\ Department of Energy,
Nuclear Physics Division, under Contract No.\ W-31-109-ENG-38.


\begin{figure}
\caption{Effect of the full electromagnetic interaction on the $pd$
         scattering observables at $E_p=3$ ($E_d=6$) MeV.
         Dashed curve: purely nuclear plus charge Coulomb contribution;
         solid curve: inclusion of full electromagnetic contribution.
         Data are from Shimizu {\it et al.}~\protect\cite{Shi95}.}
\label{fig1}
\end{figure}

\begin{figure}
\caption{Effect of the full electromagnetic interaction on the $nd$
         scattering observables at $E_n=3$ ($E_d=6$) MeV.
         Dashed curve: nuclear contribution; solid curve: inclusion
         of full electromagnetic contribution.
         Data are from Schwarz {\it et al.}~\protect\cite{Sch83} and
         McAninch {\it et al.}~\protect\cite{McA94}.}
\label{fig2}
\end{figure}

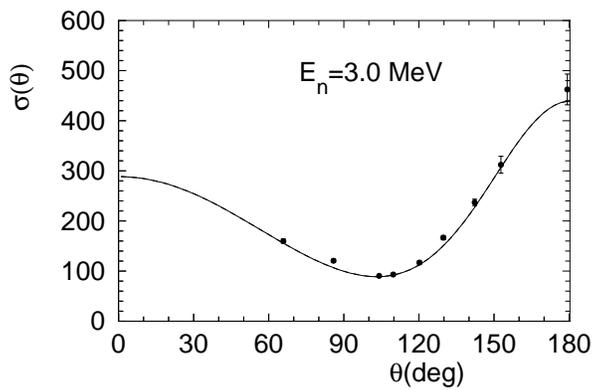
\end{document}